\documentclass[aps,prl,twocolumn,showpacs,superscriptaddress]{revtex4}
\usepackage{graphicx}
\usepackage{bm}
\usepackage{indentfirst}
\usepackage[english]{babel}
\usepackage{amssymb}
\usepackage{graphicx}

\begin{document}

\title{Asymmetric tunneling, Andreev reflection
and dynamic conductance spectra in strongly correlated metals}
\author{V.R. Shaginyan}
\email{vrshag@thd.pnpi.spb.ru} \affiliation{Petersburg Nuclear
Physics Institute, RAS, Gatchina, 188300, Russia}
\author{K.G. Popov}
\address{Komi Science Center,
Ural Division, RAS, Syktyvkar, 167982, Russia}


\begin{abstract}
Landau Fermi liquid theory predicts that the differential
conductivity between metallic point and metal is a symmetric
function of voltage bias $V$. This symmetry holds if the
particle-hole symmetry is preserved. We show that the situation can
be different when one of the two metals is a strongly correlated
one whose electronic system can be represented by a heavy fermion
liquid. When the heavy fermion liquid undergoes fermion
condensation quantum phase transition, the particle-hole symmetry
is violated making both the differential tunneling conductivity and
dynamic conductance asymmetric as a function of applied voltage.
This asymmetry can be observed when the strongly correlated metal
is either normal or superconducting. We show that at small values
of $V$ the asymmetric part of the dynamic conductance is a linear
function of $V$ and inversely proportional to the maximum value of
the gap and does not depend on temperature provided that metal is
superconducting, when it becomes normal the asymmetric part
diminishes at elevated temperatures.
\end{abstract}

\pacs{71.27.+a, 74.20.Fg, 74.25.Jb\\
{\it Key Words}: quantum phase transitions, heavy fermions,
quasiparticles, asymmetric conductivity}

\maketitle


\section{Introduction}

The unusual properties of strongly correlated electron liquid
recently observed in high-$T_c$ superconductors (HTSC) and
heavy-fermion (HF) metals are due to quantum phase transitions
taking place at their critical points. Therefore, direct
experimental studies of these transitions and critical points are
of crucial importance for understanding the physics of strongly
correlated metals. In case of HTSC such direct experimental studies
are absent since the corresponding critical points are "shadowed"
by the superconductivity. Recent experimental data on the behavior
of HF metals pour a light on the nature of these critical points
and corresponding phase transitions.  Thus, it is extremely
important to study simultaneously both the high-$T_c$
superconductivity and the anomalous behavior of HF metals. A
quantum phase transition is driven by control parameters such as
composition, pressure, density $x$ of electrons (holes), magnetic
field $H$, etc, and occurs at temperature $T=0$. QCP separates an
ordered phase generated by quantum phase transition from previously
existing disordered phase. It is expected that the universal
behavior is only observable if the heavy electron liquid in
question is very near QCP, for example, when the correlation length
is much larger than the microscopic length scales. Quantum phase
transitions of this type are quite common, and we shall label them
as conventional quantum phase transitions (CQPT). In this case, the
approach known as the Moria-Hertz-Millis theory is dealing with
some kind of magnetically ordered state of a substance and focuses
on the long wavelength low energy thermal and quantum fluctuations
on this state background. The corresponding critical state near QCP
is characterized by the complete absence of quasiparticles. It is
commonly believed that the absence of quasiparticle-like
excitations is the main cause of the non-Fermi liquid (NFL)
behavior and other types of the critical behavior in the quantum
critical region. However, the predictions of theories based on CQPT
fail to explain the strong NFL behavior observed in such HF metals
as CeCu$_{6-x}$Au$_x$, CeCoIn$_5$ and YbRh$_2$Si$_2$ near the
magnetic ordering transition or in such HF metal as $\rm
CeRu_2Si_2$ showing neither evidence of the magnetic ordering
and/or superconductivity down to the lowest temperatures nor the
LFL behavior \cite{takah}. Therefore, we conclude that the magnetic
ordering does not play the key role in the NFL behavior and can be
considered as a side effect adding specific features to the
corresponding NFL behavior.

The Landau Fermi liquid (LFL) theory rests on both the notion of
quasiparticles representing elementary excitations of a Fermi
liquid and the concept of the order parameter which characterizes
the ordered phase. Quasiparticles are appropriate excitations to
describe its low temperature thermodynamic properties. The
inability of the LFL theory to explain the experimental
observations which point to the dependence of quasiparticle
effective mass $M^*$ on temperature $T$ and applied magnetic field
$H$ may lead to the conclusion that Landau paradigm related to the
quasiparticles and order parameter fail to explain the experimental
facts on the NFL behavior of strongly correlated electrons
\cite{cust,senth}. However, other experimental facts show that
quasiparticles do exist near CQP \cite{fujim,wf,jpag} while $M^*$
depends strongly on magnetic field and temperature \cite{shag}. As
a result, we can safely conclude that the fluctuations are not
responsible for observed NFL behavior, the quasiparticles are not
suppressed by them \cite{smsk} and Landau paradigm survives also
when dealing with strongly correlated electrons \cite{shag}.
Therefore the problem of the NFL behavior can be resolved within
the LFL theory providing that quasiparticles form so-called
fermion-condensate (FC) state \cite{ks,vol} emerging behind the
fermion condensation quantum phase transition (FCQPT) \cite{ams}.
It is possible to explain the "whole bunch" of observed
thermodynamic properties of HF metals on the basis of FCQPT
allowing for existence of Landau quasiparticles down to the lowest
temperatures \cite{shag}.

The experiments on HF metals and HTSC explore mainly their
thermodynamic properties. It is highly desirable to probe the other
properties of the heavy electron liquid like quasiparticle
occupation numbers, which are not directly linked to the density of
states or to the behavior of the effective mass $M^*$. Both
scanning tunneling microscopy (STM) and point contact spectroscopy
(PCS) based on Andreev reflection (AR) \cite{andr} being sensitive
to both the density of states and quasiparticle occupation numbers
are ideal techniques to study the effects of particle-hole symmetry
violation, making the differential tunneling conductivity and
dynamic conductance to be asymmetric function of applied voltage.
This asymmetry can be observed when HF metals and HTSC are either
normal or superconducting. We note that this asymmetry is unusual
in conventional metals, especially at low temperatures. In the case
of LFL the particle-hole symmetry conserves and both the
differential tunneling conductivity and dynamic conductance are
symmetric functions of voltage bias. Thus, STM and PCS provide a
new direction in the experimental studies of the NFL behavior of
HTSC and HF metals.

In this letter we show that a particle-hole symmetry is violated
when the heavy electron liquid undergoes FCQPT. As a result, both
the differential tunneling conductivity $\sigma_d(V)$ and the
dynamic conductance $\sigma_c(V)$ become asymmetric as a function
of voltage $V$. We demonstrate that the asymmetric part is defined
by a temperature independent part of the entropy characterizing the
normal state of the electronic system of a strongly correlated
metal, and the study of the thermodynamic properties is possible by
measuring the transport characteristics of the metal. While the
application of magnetic field destroys the NFL behavior of the
heavy electron liquid and restores the above symmetry. Therefore
the measurements of asymmetric part of the conductance can be
viewed as a powerful tool to investigate the NFL behavior of
strongly correlated Fermi systems.

\section{Symmetric tunneling and FCQPT}

The tunneling current $I$ through the point contact between two
ordinary metals is proportional to the driving voltage $V$ and to
the squared modulus of the quantum mechanical transition amplitude
$t$ multiplied by the difference $N_1(0)N_2(0)(n_1(p,T)-n_2(p,T))$
\cite{harr}. Here $n(p,T)$ is the quasiparticle distribution
function (occupation number) and $N(0)$ is the density of states of
the corresponding metal. On the other hand, the wave function
calculated in WKB approximation is proportional to
$(N_1(0)N_2(0))^{-1/2}$. As a result, the density of states cancels
down and the tunneling current becomes independent of
$N_1(0)N_2(0)$ \cite{harr}.

Now we turn to a consideration of the tunneling current at low
temperatures which in the case of ordinary metals is given by
\cite{guy}

\begin{equation}\label{ui1}
I(V)=2|t|^2\int\left[n_F(\varepsilon-V)-
n_F(\varepsilon)\right]d\varepsilon.
\end{equation}

Here $n_F(\varepsilon)$ is the distribution function of ordinary
metal. We use an atomic system of units: $e=m=\hbar =1$, where $e$
and $m$ are electron charge and mass, respectively. Since
temperature is low, we approximate $n_F(\varepsilon)$ by the step
function $\theta(\varepsilon-\mu)$, $\mu$ is the chemical
potential. It follows from Eq. (\ref{ui1}) that quasiparticles with
the energy $\varepsilon$, $\mu-V\leq \varepsilon\leq \mu$,
contribute to the current. One obtains from Eq. (\ref{ui1}) that
$I(V)=a_1V$ and $\sigma_d(V)=dI/dV=a_1$, $a_1=const$. Thus, within
the LFL theory the differential tunneling conductivity
$\sigma_d(V)$ is a symmetric function of the voltage $V$. In fact,
the symmetry of $\sigma_d(V)$ holds if the particle-hole symmetry
is conserved as it is the case in the LFL theory. Therefore, the
existence of the $\sigma_d(V)$ symmetry is quite obvious and common
in the case of metal-to-metal contacts when those metals are
ordinary in their normal or superconducting states.

Now we briefly describe the heavy electron liquid with FC
\cite{shag1,shag}. When the density $x$ of the liquid approaches
some threshold value $x_{FC}$ the effective mass diverges. Beyond
$x_{\rm FC}$ the effective mass becomes negative. To avoid
physically meaningless states with $M^*<0$, the system undergoes
FCQPT with FC formation behind in the critical point $x=x_{\rm
FC}$. Therefore, behind the critical point $x_{FC}$, the step-wise
quasiparticle distribution function $n(p,T\to0)=\theta(p-p_F)$ does
not deliver the minimum to the Landau functional $E[n({\bf p})]$.
As a result, at $x<x_{FC}$ and $T=0$ the quasiparticle distribution
is determined by the standard equation for the minimum of a
functional \cite{ks}
\begin{equation}\label{ui2}
\frac{\delta E[n({\bf p})]}{\delta n({\bf p},T=0)}=\varepsilon({\bf
p})=\mu; \,p_i\leq p\leq p_f.
\end{equation}
Equation (\ref{ui2}) determines the quasiparticle distribution
function $n_0({\bf p})$ minimizing the ground state energy $E\equiv
E[n({\bf p})]$. Being determined by Eq. (\ref{ui2}), the function
$n_0({\bf p})$ does not coincide with the step function
$\theta(p-p_F)$ in the region $(p_f-p_i)$, so that $0<n_0({\bf
p})<1$, while outside the region it coincides with $\theta(p-p_F)$.
It follows from Eq. (\ref{ui2}) that the single particle spectrum
is completely flat in this region. Such a state was called the
state with FC since the quasiparticles in the range $p_f-p_i$ of
momentum space are confined to the chemical potential $\mu$. The
possible solution $n_0({\bf p})$ of Eq. (\ref{ui2}) and the
corresponding single particle spectrum $\varepsilon({\bf p})$ are
shown in Fig. 1. It is seen from Fig. 1 that the particle-hole
symmetry is not supported by FC since $n_0({\bf p})$ does not
evolve from the Fermi-Dirac distribution function and we can expect
that the conductivity possesses an asymmetric part.
\begin{figure}[!ht]
\begin{center}
\includegraphics[width=0.47\textwidth]{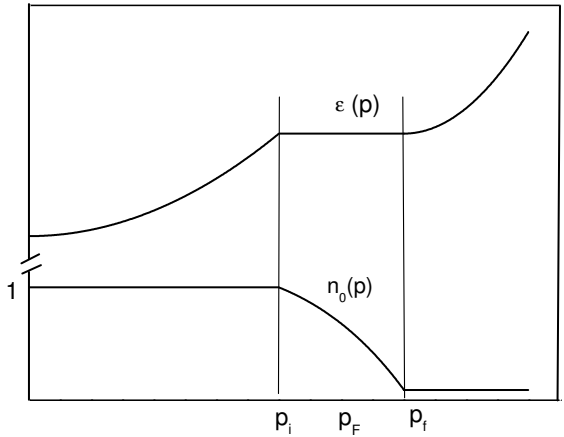}
\end{center}
\caption{The quasiparticle distribution function $n_0(p)$ and energy
$\varepsilon(p)$. Since $n_0(p)$ is the solution of Eq. (\ref{ui2})
it implies $n_0(p<p_i)=1$, $n_0(p_i<p<p_f)<1$ and $n_0(p>p_f)=0$,
while $\varepsilon(p_i<p<p_f)=\mu$. The Fermi momentum $p_F$ obeys
the condition $p_1<p_F<p_f$.} \label{Fig1}
\end{figure}
One can show that the relevant order parameter $\kappa({\bf
p})=\sqrt{n_0({\bf p})(1-n_0({\bf p}))}$ is the order parameter of
the superconducting state with the infinitely small superconducting
gap $\Delta$ and with the entropy $S=0$ \cite{smsk,ams}. Hence this
state cannot exist at any finite temperatures and is driven by the
parameter $x$: at $x>x_{FC}$ the system is on the disordered side
of FCQPT; at $x=x_{FC}$, Eq. (\ref{ui2}) possesses the non-trivial
solutions $n_0({\bf p})$ with $p_i=p_F=p_f$; at $x<x_{FC}$, the
system is on the ordered side. At $T>0$, $n({\bf p},T)$ is given by
the Fermi-Dirac distribution function
\begin{equation}\label{ui3}
n({\bf p},T)=\left\{ 1+\exp
\left[ \frac{(\varepsilon({\bf p},T)-\mu)}T \right] \right\} ^{-1},
\end{equation}
where $\varepsilon({\bf p},T)$ is the single-particle spectrum,
determined by Eq. (\ref{ui2}) and $\mu$ is a chemical potential.
Equation (\ref{ui3}) can be recast as
\begin{equation}\label{ui4}
\varepsilon ({\bf p},T)-\mu (T)=T\ln \frac{1-n({\bf p},T)}{n({\bf
p},T)}.
\end{equation}
As $T\to 0$, the logarithm on the right hand side of Eq.
(\ref{ui4}) is finite when $p\in (p_f-p_i)$ and $n({\bf p},T)\simeq
n_0({\bf p})$ so that $T\ln(...)\to 0$, and we again arrive at Eq.
(\ref{ui2}). Near the Fermi level the single particle spectrum can
be approximated as
\begin{equation} \label{ui5} \varepsilon(p\simeq p_F,T)-\mu\simeq
\frac{p_F(p-p_F)}{M^*(T)}. \end{equation} It follows from Eq.
(\ref{ui3}) that $n(p,T\to 0)\to \theta(p-p_F)$ if $M^*$ is finite
at $T\to 0$. At low temperatures, as it is seen from Eq.
(\ref{ui5}), the effective mass diverges as \cite{ams}
\begin{equation} \label{ui6} M^*(T)\simeq p_F\frac{p_f-p_i}{4T}.
\end{equation}
At $T\ll T_f$, Eq. (\ref{ui5}) is valid and describes the
quasiparticles with the energy $\varepsilon $ and the distribution
function $n_0(p)$. Here $T_f$ is the characteristic temperature
where the influence of FCQPT becomes negligible \cite{ams}. The
energy $\varepsilon$ belongs to the interval
\begin{equation}
\label{ui7} \mu-2T\leq \varepsilon \leq\mu+2T.
\end{equation}
When the heavy electron liquid becomes superconducting the
effective mass is temperature independent at $T\leq T^*_c$ and
given by the equation \cite{ams,ashag}
\begin{equation} \label{ui61} M^*(T)\simeq
p_F\frac{p_f-p_i}{2\Delta_1}, \end{equation} where $T^*_c$ is the
transition temperature at which the gap vanishes and $\Delta_1$ is
the maximum value of the superconducting gap at $T=0$.

\section{Asymmetric conductance in HF metals and HTSC}

In the case of the heavy electron liquid with FC, the tunneling
current is of the form \cite{tun}
\begin{equation} \label{ui8}
I(V)=\int\left[n(\varepsilon-V,T)-
n_F(\varepsilon,T)\right]d\varepsilon.
\end{equation}
Here we have replaced the distribution function of ordinary metal
by $n(\varepsilon,T)$ so that $n(\varepsilon(p),T\to0)\to
n_0(\varepsilon(p))$ where $n_0(\varepsilon(p))$ is the solution of
Eq. (\ref{ui2}) and also normalized the transition amplitude
$|t|^2=1$. The differential tunneling conductivity,
$\sigma_d(V)=\partial I/\partial V$, is given by
\begin{equation}\label{ui9}
\sigma_d=\frac{1}{T}\int n(\varepsilon(z)-V,T)
(1-n(\varepsilon(z)-V,T)) \frac{\partial \varepsilon(z,T)}{\partial
z}dz,
\end{equation}
where $z=p/p_F$. We take dimensionless momentum $z$, instead of
energy $\varepsilon $, as an independent variable, since the
distribution function $n$ is a continuous function of $z$ rather
than of $\varepsilon$ as it seen from Fig. 1. Indeed, the energy
$\varepsilon$ is a constant in the range $p_i-p_f$ while the
distribution function varies in this range. It follows from Eq.
(\ref{ui9}) that the asymmetric part $\Delta
\sigma_d(V)=(\sigma_d(V)-\sigma_d(-V))/2$ of the differential
conductivity is of the form
\begin{eqnarray}
\Delta\sigma_d(V)&=&\frac{1}{2}\int\frac{\partial n(z,T)}{\partial
z} \frac{\alpha(1-\alpha^2)}
{[n(z,T)+\alpha[1-n(z,T)]^2} \nonumber\\
&\times&\frac{1-2n(z,T)}{[\alpha
n(z,T)+[1-n(z,T)]]^2}dz,\label{ui10}
\end{eqnarray}
with $\alpha=\exp(-V/T)$. It is worth noting that according to Eq.
(\ref{ui10}) we have $\Delta \sigma_d(V)=0$ if the considered HF
metal is replaced by an ordinary metal. Indeed, the effective mass
is finite at $T\to 0$ so that the integrand becomes an odd function
of $x=z-1$, while the limits of integration can be taken
$-\infty,\infty$ since the integrand behaves like $\exp(-|x|)$ at
large $|x|$. On the other hand, the integrand is no longer an odd
function if the particle-hole symmetry is violated. As it is seen
from Fig. 1, there are no reasons to expect that a Fermi liquid
with FC conserves this symmetry. Thus, we conclude that the
differential conductivity becomes an asymmetric function of the
applied voltage for HF liquid with FC.

To estimate $\Delta \sigma_d(V)$, we observe that it is zero when
$V=0$ as it should be and follows from Eq. (\ref{ui10}) as well. It
is seen from Eq. (\ref{ui10}) that at low voltage $V$ the
asymmetric part behaves as $\Delta \sigma_d(V)\propto V$. Then, the
natural scale to measure the voltage is $2T$, as it is seen from
Eq. (\ref{ui7}). In fact, the asymmetric part is proportional to
$(p_f-p_i)/p_F$. As a result, we obtain
\begin{equation} \label{ui11}
\Delta \sigma_d(V)\simeq
c\left(\frac{V}{2T}\right)\frac{p_f-p_i}{p_F} \simeq
c\frac{V}{2T}\frac{S_0}{x}.
\end{equation}
Here, $x=p_F^3/(3\pi^2)$ is a density of the heavy electron liquid,
$c$ is a constant of the order of unity, and $S_0$ is the
temperature independent part of an entropy $S[n({\bf p},T)]$ which
is given by the familiar expression
\begin{eqnarray}
S[n({\bf p},T)]&=&-2\int[n({\bf p},T)\ln n({\bf p},T)+(1-n({\bf p},T))\nonumber\\
&\times&\ln(1-n({\bf p},T))]\frac{d{\bf p}}{(2\pi)^3}.\label{ui12}
\end{eqnarray}
Inserting the solutions $n_0({\bf p})$ into Eq. (\ref{ui12}) we
obtain that the entropy contains a temperature independent part
$S_0/x\sim (p_f-p_i)/p_F$. At $T<T_f$, it can be approximated as
\cite{alp} \begin{equation} S(T) \simeq
S_0+a\sqrt{\frac{T}{T_f}},\label{entr1}\end{equation} here $a$ is a
constant. Thus, the ordered state existing at $T=0$ is separated
from the disordered state by first order phase transition. Due to
this first order phase transition, both at the FCQPT point and
behind it there are no critical fluctuations accompanying second
order phase transitions and suppressing the quasiparticles. As a
result, the quasiparticles survive and define the thermodynamic and
transport properties of HF systems. This is in agreement with
recent facts obtained in measurements on $\rm CeCoIn_5$
\cite{jpag}. This means that asymmetric conductivity measurements
can provide a valuable information about the entropy $S_0$ which
determines the divergence of the Gr\"uneisen ratio $\Gamma(T)$ as
well since, as it is seen from Eq. (\ref{entr1}), $\Gamma(T)\propto
S_0/\sqrt{T}$ \cite{alp}. So, we may conclude that STM and PCS
techniques can give an experimental evidence about the NFL behavior
of strongly correlated fermion systems. The constant $c$ can be
evaluated using the analytically solvable models. For example,
calculations of $c$ within a simple model, when Landau functional
$E[n(p)]$ is of the form \cite{ksk}
\begin{eqnarray} \label{ui13} E[n(p)]&=&\int \frac{{\bf p}^2}{2m}\frac{d{\bf
p}}{(2\pi)^3}+\frac{1}{2}\int V({\bf p}_1-{\bf p}_2)n({\bf
p}_1)n({\bf p}_2)\nonumber \\&\times&\frac{d{\bf p}_1d{\bf
p}_2}{(2\pi)^6},
\end{eqnarray}
with inter-particle interaction
\begin{equation}\label{ui14}
V({\bf p})=g_0\frac{\exp(-\beta_0|{\bf p}|)}{|{\bf p}|},
\end{equation}
gives $c\simeq 1/2$. It follows from Eq. (\ref{ui11}), that when
$V\simeq 2T$ and FC occupies a noticeable part of Fermi volume,
$(p_f-p_i)/p_F\simeq 1$, so that the asymmetric part becomes
comparable with differential conductivity, $\Delta \sigma_d(V)\sim
\sigma_d(V)\sim \sigma_c(V)$.

The asymmetric behavior of the conductivity can be observed in
measurements on both high-$T_c$ metals in their normal state and
the heavy fermion metals, for example, such as $\rm CeCoIn_5$ and
YbRh$_2$(Si$_{0.95}$Ge$_{0.05}$)$_2$ which are expected to have
undergone FCQPT. Indeed, the measurements on these metals have
shown that the Gr\"uneisen ratio diverges \cite{cust,geg,oes}. In
that case, the electronic systems of these metals are to undergo
FCQPT so that the temperature independent part $S_0$ of the entropy
becomes finite \cite{alp}. As a result, the asymmetric part of the
conductance $\Delta\sigma_d(V)$ becomes finite \cite{tun}. We note
that at sufficiently low temperatures, the application of magnetic
field restores the LFL behavior making the asymmetry of the
tunneling conductivity vanish \cite{tun}. Therefore, the
measurements have to be carried out when the corresponding HF metal
demonstrates the NFL behavior.

Point contact spectroscopy has recently been used to investigate
the HF metal $\rm CeCoIn_5$. The dynamic conductance spectra shown
in Fig. 2 have been obtained and the asymmetric character of
conductance has been observed in  $\rm CeCoIn_5$ both in its
superconducting and NFL normal states \cite{park}. Fig. 2 shows the
conductance $dI/dV$ as a function of voltage $V$. The finite and
even enhanced subgap conductance seen in Fig. 2 as well as that
below the critical temperature $T_c$ of superconducting phase
transition arises from Andreev reflection, see e.g. \cite{guy}. It
is seen from Fig. 2 that the asymmetric conductivity develops at
about 45 K, that is well above the critical temperature $T_c=2.3$ K
of superconducting phase transition. The asymmetry becomes more
pronounced with decreasing temperature down to $T_c$. Then it
remains almost the same at least down to $0.4$ K \cite{park}.

\begin{figure} [! ht]
\begin{center}
\includegraphics [width=0.47\textwidth] {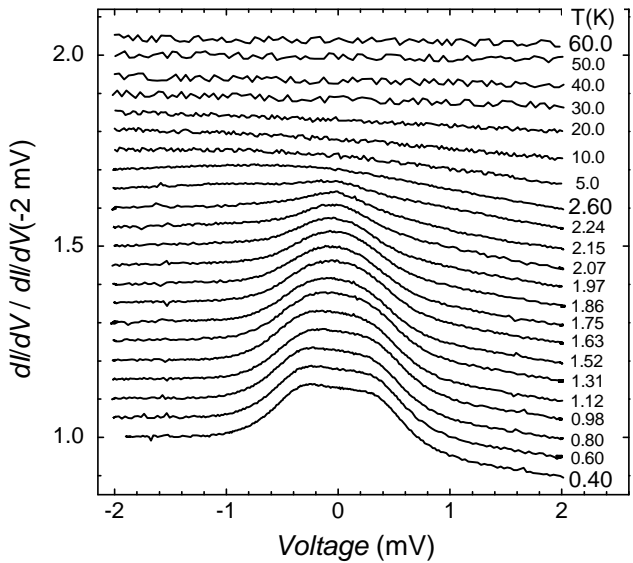}
\end {center}
\caption{ Dynamic conductance spectra $\sigma_c(v)=dI/dV$ measured
with point contact (Au/CeCoIn$_5$) over wide temperature range
\cite{park}. Curves $\sigma_c(v)$ are shifted  vertically by 0.05
for clarity and normalized by the conductance at -2 mV. The
asymmetry is seen to develop starting at $45$ K and becomes
stronger with decreasing temperature \cite{park}.} \label {Fig2}
\end{figure}

In Fig. 3, we plot the results of our calculations (based on the
functional (\ref{ui13})) of the asymmetric part $\Delta
\sigma_d(V)$ of the conductance $\sigma_c(V)$ as a function of
voltage when the heavy electron liquid is in its normal state.

\begin{figure}[!ht]
\begin{center}
\includegraphics[width=0.47\textwidth] {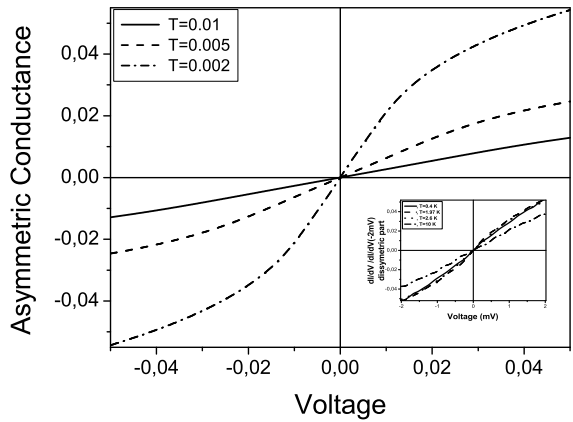}
\end{center}
\caption{The asymmetric conductance $\Delta \sigma_d(V)$ vs.
normalized voltage calculated for three values of the normalized
temperature. The values of the normalized temperature are shown in
the upper left corner. The asymmetric part of the conductance
extracted from data of Fig. 2 are shown in the inset.} \label{Fig3}
\end{figure}

We normalize both the voltage $V$ and temperature $T$ by the Fermi
energy $\varepsilon^0_F$. We use the dimensionless coupling constant
$g=(g_0 M)/(2\pi^2)$ and $\beta=\beta_0 p_F$. FCQPT takes place when
the parameters reach their critical values, $\beta=b_c$ and $g=g_c$.
In the considered case, we take $\beta=3$ (the corresponding
$g=g_c=6.7167$) and $g=8$ so that $(p_f-p_i)/p_F\sim 0.1$. It is
seen from Fig. 2 that asymmetric conductance vanishes at elevated
temperatures when $T\to T_f$. The asymmetric part of the conductance
$\Delta \sigma_d(V)$, extracted from the data of Fig. 2, is reported
in the inset to Fig. 3. It is seen that our calculations are in good
qualitative agreement with experimental data. Although we were not
aiming to attain the detailed quantitative agreement with the data
within our simple model, a better agreement can be achieved by
taking the appropriate Fermi energy value. The corresponding results
will be published elsewhere.

The asymmetric conductivity $\Delta \sigma_d(V)$ can also be
observed when both HTSC and HF metals in question go from normal to
superconducting phase. The reason now is that $n_0({\bf p})$ is
again responsible for the asymmetric part of the differential
conductivity measured by both STM and PSC. As it was shown in Ref.
\cite{shag}, the function $n_0({\bf p})$ is not appreciably
disturbed by the superconductive pairing interaction which is
relatively weak as compared to Landau interaction forming the
distribution function $n_0({\bf p})$. Therefore, the asymmetric
conductance remains approximately the same below $T_c$. This result
is in good agreement with experimental facts as it is seen from the
lower inset of Fig. 3. We also conclude that Andreev reflection can
be considered as a useful effect when studying the asymmetric
conductance and the NFL behavior. When calculating the tunneling
conductance measured by scanning tunneling microscopy, we have to
take into account that
\begin{equation} \label{ui15} N_s(E)=
N(\varepsilon-\mu)\frac{E}{\sqrt{E^2-\Delta^2}}
\end{equation}
comes now into the play since the density of states $N_s(E)$ of the
superconducting metal is zero in the gap i.e. when $E\leq
|\Delta|$. Here, $\Delta$ is the superconducting gap at $T=0$, $E$
is the quasiparticle energy in the superconducting state, related
to the normal state quasiparticle energy as
$\varepsilon-\mu=\sqrt{E^2-\Delta^2}$. It follows from Eq.
(\ref{ui15}) that the tunneling conductance can be asymmetric if
the density of states $N(\varepsilon-\mu)$ is asymmetric with
respect to the Fermi level \cite{pand} as it is in the case of
Fermi systems with FC.
\begin{figure} [! ht]
\begin{center}
\includegraphics [width=0.47\textwidth] {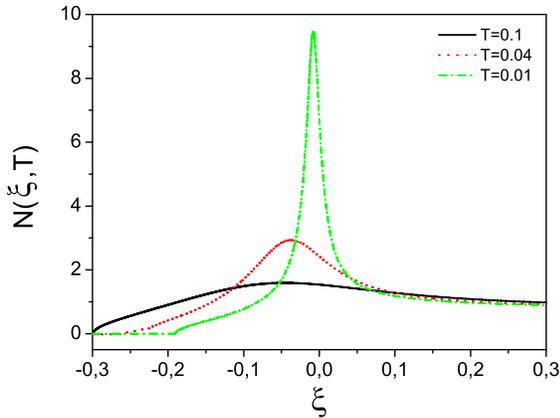}
\end {center}
\caption{The density of states $N(\xi,T)$ as a function of
$\xi=(\varepsilon-\mu)/\mu$ calculated for the three values of the
normalized temperature $T$. The values of the normalized
temperature are shown in the upper right corner.} \label {Fig4}
\end{figure}
In Fig. 4, the results of our calculations (based on the functional
(\ref{ui13}) with the same parameters when calculating the
asymmetrical conductance shown in Fig. 3) of the density of states
$N(\xi,T)$ as a function of normalized binding energy
$\xi=(\varepsilon-\mu)/\mu$ are shown. It is seen from Fig. 4 that
$N(\xi,T)$ is strongly asymmetric with respect to the Fermi level,
or $\mu$. As a result, at the Fermi level the first derivative of
$N(\xi,T)$ with respect to $\xi$ is not zero, and at small values
of $\xi$ $N(\xi,T)$ can be approximated as $N(\xi,T)\simeq
a_0+a_1\xi$. The coefficient $a_0$ does not contribute to the
asymmetric conductance which is obviously defined by $a_1$. The
coefficient $a_1$ has to be proportional to the effective mass
related to $T$ by Eq. (\ref{ui6}). Now the system in its
superconducting state and the effective mass is given by Eq.
(\ref{ui61}) so that the density of states is determined by
$\Delta_1$ while the values of the normalized temperature shown in
the upper right corner of Fig. 4 are defined as $2T\simeq
\Delta_1/\varepsilon_F.$ So we have $\Delta \sigma_d(V)\sim
VS_0/x\Delta_1$ since $(p_f-p_i)/p_F\simeq S_0/x$, while $E=V$ and
$\xi=\sqrt{E^2-\Delta^2}$. It is instructive to adjust Eq.
(\ref{ui11}) for the case of superconducting HF metal, multiplying
the right hand side of this expression by $N_s(E)$ and replacing
the quasiparticle energy $\varepsilon-\mu$ by $\sqrt{E^2-\Delta^2}$
with $E$ being substituted by the voltage $V$. As a result, Eq.
(\ref{ui11}) can be again presented in the form
\begin{equation} \Delta \sigma_d(V)\simeq
\frac{V}{\Delta}\frac{\sqrt{V^2-\Delta^2}} {\sqrt{V^2-\Delta^2}}
\frac{p_f-p_i}{p_F} \simeq \frac{V}{\Delta_1}\frac{S_0}{x}.
\label{ui16}\end{equation} We note that the entropy $S_0$ on right
hand side of Eq. (\ref{ui16}) is a temperature independent part of
the normal (rather then superconducting) state entropy. Thus, as it
follows from Eq. (\ref{ui16}), the entropy $S_0$ characterizing the
normal state can be evaluated measuring the asymmetric part of
tunneling conductance of the superconducting state. We note that
the scale $2T$ entering Eq. (\ref{ui11}) is replaced by the scale
$\Delta_1$ in Eq. (\ref{ui16}). In the same way, as Eq.
(\ref{ui11}) is valid up to $V\sim 2T$, Eq. (\ref{ui16}) is valid
up to $V\sim \Delta_1$ and up to temperatures $T\leq T_c$. It is
seen from Eq. (\ref{ui16}) that the asymmetric part of the
differential tunneling conductivity becomes as large as the entire
differential tunneling conductivity at $V\sim \Delta_1$ under the
condition that FC occupies a large part of the Fermi volume,
$(p_f-p_i)/p_F\simeq 1$. In the case of a $d$-wave (or other
unconventional) pairing, the right hand side of Eq. (\ref{ui16})
has to be additionally integrated over the corresponding gap
function. In our case this procedure is trivial, and $\Delta
\sigma_d(V)$ becomes finite even at $V\leq \Delta_1$. In the case
of PCS due to Andreev reflection the asymmetrical conductance can
be observed even at $V\to0$ and in the case of $s$-wave pairing.
\begin{figure}[!ht]
\begin{center}
\includegraphics[width=0.47\textwidth]{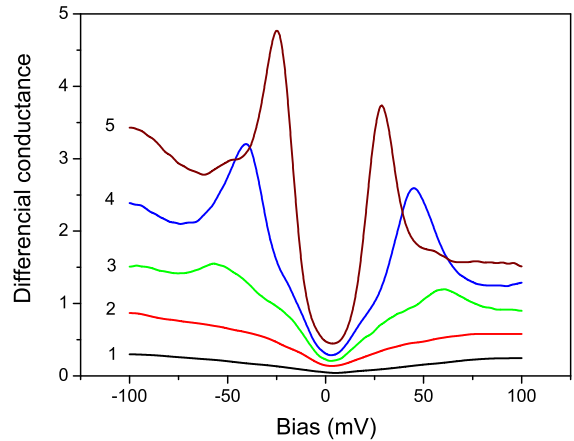}
\end{center}
\caption{Spatial variation of the tunneling differential
conductance $\Delta \sigma_d(V)$ spectra measured in $\rm
Bi_2Sr_2CaCu_2O_{8+x}$. Curves 1 and 2 are taken at the positions
where the integrated LDOS is very small. The low differential
conductance and the absence of a superconducting gap are indicative
for insulating behavior. Curve 3 is for a large gap 65 meV with
small coherence peaks. The integrated value of the LDOS at the
position for curve 3 is small but larger than those in curves 1 and
2. Curve 4 is for a gap of 40 meV, which is close to the mean value
of the gap function. Curve 5, taken at the position with the
highest integrated LDOS, is for the smallest gap of 25 meV with two
very sharp coherence peaks \cite{pan}.} \label{Fig5}
\end{figure}
The tunneling differential conductances $\Delta \sigma_d(V)$
depending on the local density of states (LDOS) was obtained in
measurements on $\rm Bi_2Sr_2CaCu_2O_{8+x}$ at low temperatures
\cite{pan} and is shown in Fig. 5. The presence of an electronic
inhomogeneity in $\rm Bi_2Sr_2CaCu_2O_{8+x}$ has recently been
discovered in the scanning tunneling microscopy and spectroscopy
experiments. This inhomogeneity is manifested as spatial variation
in LDOS spectrum, in the low-energy spectral weight, and in the
magnitude of the superconducting energy gap. The inhomogeneity
observed in the integrated LDOS is not due to impurities, but
rather is intrinsic characteristic of a substance. The observations
allow to relate the magnitude of the integrated LDOS to the local
oxygen concentration. The curves shown in Fig. 5 can be viewed as
those corresponding to HTSC with different oxygen concentrations or
corresponding to HTSC with different values of $\Delta_1$ measured
at the same $T$ thus permitting to study the asymmetric conductance
as a function of $\Delta_1$. It is seen from Fig. 5 that the
tunneling differential conductivity is strongly asymmetric in
superconducting state of the $\rm Bi_2Sr_2CaCu_2O_{8+x}$ compound.
Fig. 6 presents the asymmetric parts $\Delta\sigma_d(V)$ of the
tunneling differential conductance spectra extracted from data
displayed in Fig. 5. It is seen that at small values of the bias
voltage $V$ and in accord with Eq. (\ref{ui16}),
$\Delta\sigma_d(V)$ is a linear function of $V$ and the coefficient
$a_1$ and the slope of the lines displayed in Fig. 6 are inversely
proportional to the gap $\Delta_1$, see Fig. 5.
\begin{figure} [! ht]
\begin{center}
\includegraphics [width=0.47\textwidth] {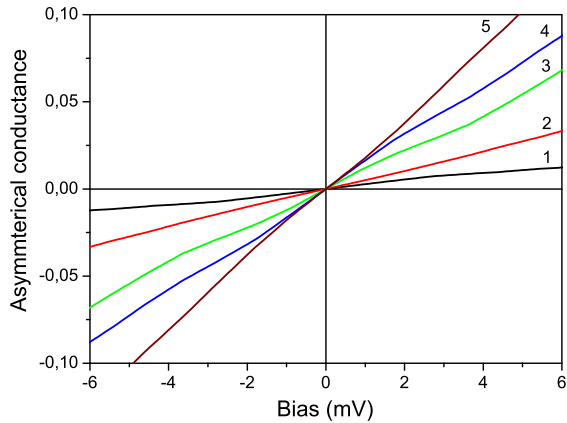}
\end {center}
\caption{The asymmetric parts $\Delta\sigma_d(V)$ of the tunneling
differential conductance spectra measured in $\rm
Bi_2Sr_2CaCu_2O_{8+x}$ and extracted from data of Fig. 5 are shown
as functions of the bias (mV). The number of the curve representing
the asymmetric part correspond to the number of the curve displayed
in Fig. 4 from which this asymmetric part was extracted.} \label
{Fig6}
\end{figure}
Temperature dependence of the asymmetric parts $\Delta\sigma_d(V)$
of point contact spectra on $\rm
YBa_2O_{7-x}/La_{0.7}Ca_{0.3}MnO_3$ bilayers with $T_c\simeq 30$ K
shows that the asymmetric conductivity remains constant up to
temperatures of about $T_c$ and persists up to temperatures well
above $T_c$, see Fig. 7. It is also seen from Fig. 7 that at small
values of the voltage $V$ the asymmetrical part is a linear
function of $V$ and starts to diminish at $T\geq T_c$. This
behavior permits to study the asymmetric conductance as a function
of temperature and is in good agreement with the behavior described
by Eqs. (\ref{ui11}) and (\ref{ui16}). Therefore, we can conclude
that the asymmetric part of the conductances described by Eqs.
(\ref{ui11}) and (\ref{ui16}) are in good qualitative agreement
with the experimental facts displayed in Figures 2, 3, 5, 6 and 7,
while the asymmetry starts from FCQPT yielding the FC state with
particle-hole symmetry violation.
\begin{figure} [! ht]
\begin{center}
\includegraphics [width=0.47\textwidth] {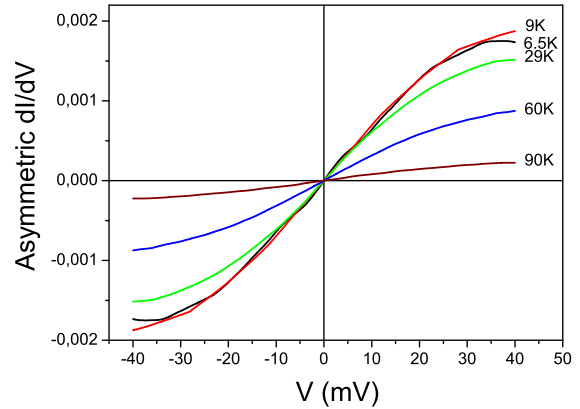}
\end {center}
\caption{Temperature dependence of the asymmetric parts
$\Delta\sigma_d(V)$ of conductance spectra extracted from
measurements on  $\rm YBa_2O_{7-x}/La_{0.7}Ca_{0.3}MnO_3$ bilayers
of the conductance spectra at different temperatures shown in the
upper right part of the figure \cite{samanta}.} \label{Fig7}
\end{figure}
\section{Conclusions}
We have shown that scanning tunneling microscopy and point contact
spectroscopy being sensitive to both the density of states and the
quasiparticle occupation numbers are ideal techniques for studying
the effects related to a violation of the particle-hole symmetry.
Above effects make the differential tunneling conductivity and
dynamic conductance to be asymmetric function of applied voltage.
The asymmetry appears as soon as FCQPT emerges and is closely
related to the violation of a particle-hole symmetry. We have
demonstrated that the asymmetric part of the conductance can be
observed when HF metals and high-$T_c$ superconductors are normal
and/or superconducting. We have shown that at small values of the
voltage bias the asymmetric part is a linear function of the
voltage and inversely proportional to the maximum value of the gap
and does not depend on temperature provided that metal is
superconducting, when it becomes normal the asymmetric part
diminishes at elevated temperatures. Our theoretical results are in
good agreement with available experimental data while it proved out
to be very useful to explore the asymmetric part of conductivity
rather then the conductivity. Since in pure LFL case the
particle-hole symmetry is conserved and both the differential
tunneling conductivity and dynamic conductance are symmetric
functions of voltage bias, the measurements of asymmetric part of
the conductance can be viewed as a powerful tool to investigate the
NFL behavior of strongly correlated Fermi systems. We have derived
that the asymmetric part is defined by the temperature independent
part of the entropy characterizing the normal state of the
electronic system of a strongly correlated metal. As a result, it
became possible to study the thermodynamic properties measuring the
transport characteristics of a substance. Thus, STM and PCS provide
a new and very useful direction in the experimental studies of the
NFL behavior of HTSC and HF metals.

Finally, our consideration of the conductance asymmetry observed in
both HF metals and high-$T_c$ superconductors and described within
the same framework suggests that FCQPT is intrinsic to strongly
correlated substances and can be viewed as the universal cause of
the non-Fermi liquid behavior.

This work was supported in part by RFBR, project No. 05-02-16085.

\end{document}